\documentclass{PoS}
\usepackage{txfonts,graphicx,subfigure,natbib}

\title{Gamma-Ray Bursts Overview}

\ShortTitle{Gamma-Ray Bursts Overview}

\author{\speaker{B.~McBreen}\\
        UCD School of Physics, University College Dublin, Belfield, Dublin 4 \\
        E-mail: \email{brian.mcbreen@ucd.ie}}

\author{S.~Foley\\
        UCD School of Physics, University College Dublin, Belfield, Dublin 4 \\
        E-mail: \email{suzanne.foley2@ucd.ie}}

\author{L.~Hanlon\\
	UCD School of Physics, University College Dublin, Belfield, Dublin 4 \\
	E-mail: \email{lorraine.hanlon@ucd.ie}}

\abstract{It is now more than 40 years since the discovery of gamma-ray bursts (GRBs) and in the last two decades there has been major progress in the observations of bursts, the afterglows and their host galaxies. This recent progress has been fueled by the ability of gamma-ray telescopes to quickly localise GRBs and the rapid follow-up observations with multi-wavelength instruments in space and on the ground. A total of 674 GRBs have been localised to date using the coded aperture masks of the four gamma-ray missions, \textit{BeppoSAX}, \textit{HETE II}, \textit{INTEGRAL} and \textit{Swift}. As a result there are now high quality observations of more than 100 GRBs, including afterglows and host galaxies, revealing the richness and progress in this field. The observations of GRBs cover more than 20 orders of magnitude in energy, from $10^{-5}$\,eV to $10^{15}$ eV and also in two non-electromagnetic channels, neutrinos and gravitational waves. However the continuation of progress relies on space based instruments to detect and rapidly localise GRBs and distribute the coordinates.}

\FullConference{The Extreme sky: Sampling the Universe above 10 keV - extremesky2009,\\
		October 13-17, 2009\\
		Otranto (Lecce) Italy}

\begin{document}

\section{Introduction}

Gamma-ray bursts (GRBs) are powerful flashes of gamma-rays appearing in the sky with durations ranging from a fraction of a second to over 1000 seconds. The isotropic equivalent energy radiated is between $10^{48}$ and $10^{55}$ ergs. The gamma-rays are thought to originate from a highly relativistic outflow with Lorentz factor $\Gamma$\,$>$100 that is ejected from a central source. The content of the jet, the level of magnetisation and the mechanism generating the gamma-rays are highly uncertain.  In this short overview we cover a small number of topics, such as durations, spectral lags, high-energy emission and afterglows, and refer the reader to much longer recent review articles \cite{gehrels:2009,meszaros:2006,nakar:2007,piran:2004,zhang:2004} and the excellent book on GRBs by Vedrenne and Atteia (2009) \cite{vedrenne:2009}.


It is interesting to compare the results obtained with recent missions between July 1996 and December 2009 (Table~\ref{table:ags}). There is a total of 674 GRBs detected by these 4 missions and most of them are better localised with X-ray measurements of the afterglow. The X-ray, optical and
radio afterglow detections are listed in Table~\ref{table:ags} and the data are taken from the webpage maintained by Jochen
Greiner\footnote{http://www.mpe.mpg.de/$\sim$jcg/grbgen.html.}. It should be noted the \textit{Swift} has detected the largest number of GRBs and has the highest percentage of detections of X-ray and optical afterglows because it has an onboard X-ray detector and an optical/UV telescope that immediately slews to the position of the burst and provides accurate coordinates for ground-based observations. \textit{Swift} has set the standard in this field. New GRB space missions should have either an X-ray telescope or infrared telescope on board to localise the GRB and enable rapid follow-up observations with other space and ground-based telescopes. 

\begin{table}[b]
\begin{center}
\caption{Afterglow detections for GRBs localised by recent $\gamma$-ray missions
between July 1996 and December 2009.}
\label{table:ags}
\begin{tabular}{@{}l c c c c @{}}
\hline\hline
& \textbf{\textit{BeppoSAX}} & \textbf{HETE II} & \textbf{\textit{INTEGRAL}} &
\textbf{\textit{Swift}} \\
\hline
\textbf{GRBs} & 55 & 79 & \textbf{66} & 474 \\
\textbf{X-ray} & 31 & 19 & \textbf{34} & 397 \\
\textbf{Optical} & 17 & 30 & \textbf{21} & 247 \\
\textbf{Radio} & 11 & 8 & \textbf{8} & 34 \\
\hline
\end{tabular}
\end{center}
\end{table}

\section{Durations and Redshifts}

\begin{figure}
\resizebox{\textwidth}{0.35\textheight}{\includegraphics{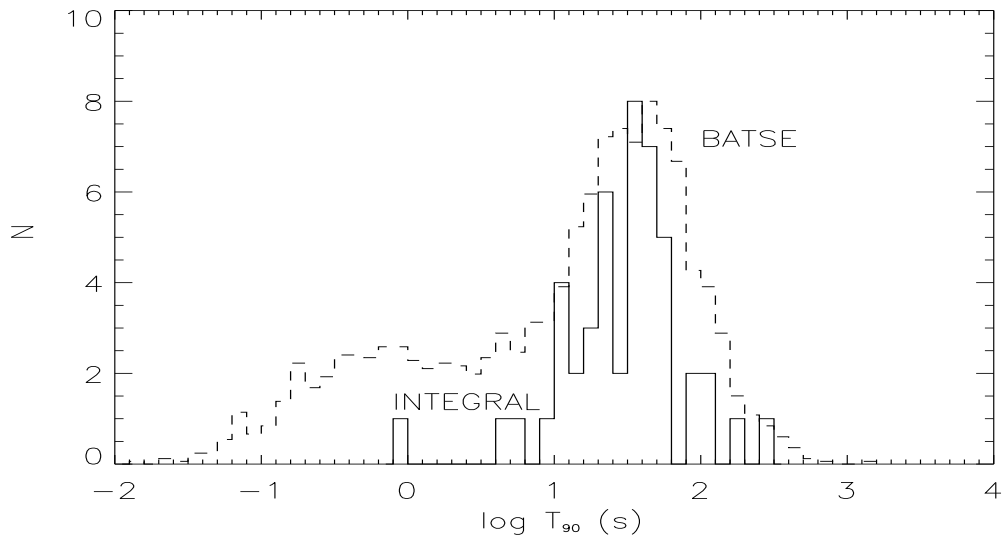}}
\caption{$T_{90}$ distribution of BATSE GRBs (dashed line) in
  comparison to that of INTEGRAL (solid line) \cite{foley:2008}. The BATSE distribution is normalised to
  the peak of the \textit{INTEGRAL} distribution for clarity. The BATSE
  data for 2041 GRBs is taken from the Current Catalog at
  http://www.batse.msfc.nasa.gov/batse/grb/catalog/current. There are two coded-mask
$\gamma$-ray instruments on board \textit{INTEGRAL}, namely
IBIS~\cite{ubertini:2003} and SPI~\cite{vedrenne:2003},
optimised for high-resolution imaging and spectroscopy of the $\gamma$-ray sky, respectively. IBIS/ISGRI has 16,384 CdTe detectors, located 3.4\,m from a tungsten mask which projects a shadowgram on the detector plane.}
\label{fig:t90s}
\end{figure}

The durations of GRBs are defined by the time during which the middle 50\% ($T_{50}$) or 90\% ($T_{90}$) of the counts are above background. It is found that there are two classes of GRBs separated at a $T_{90}$ of about 2\,s \cite{kouveliotou:1993} and are called short and long as shown in Fig.~\ref{fig:t90s}. This classification is supported by the spectral analysis where short/long GRBs are spectrally hard/soft respectively. It is often speculated that the two types of GRBs have different origins with long GRBs originating in black hole (BH) formation in supernova explosions and short GRBs in the merger of compact objects. The integrated gamma-ray profiles of some long GRBs support not only the BH formation but also the spin up and down of the newly formed BH \cite{mcbreen:2002}. The redshifts (z) have been measured for about 150 GRBs and the z distribution extends to a record high of $z=8.2$ for GRB 090423 \cite{salvaterra:2009}. \textit{Swift} detects GRBs at an average redshift of $\sim$2.5 which is larger than the pre-Swift value of  $\sim$1.2 because of the higher sensitivity of \textit{Swift} \cite{gehrels:2009}. Only GRB\,070610 has been identified as the galactic transient SWIFT J195509+261406 because of the flaring optical and X-ray afterglow \cite{castro-tirado:2008,stefanescu:2008,meehan:2010}. This is ironic given the extended debate over the galactic/extragalactic origin of GRBs.

\section{Spectral Lag}

\begin{figure}
\centering
\resizebox{0.8\hsize}{!}{\includegraphics[width=0.7\textheight,angle=270]{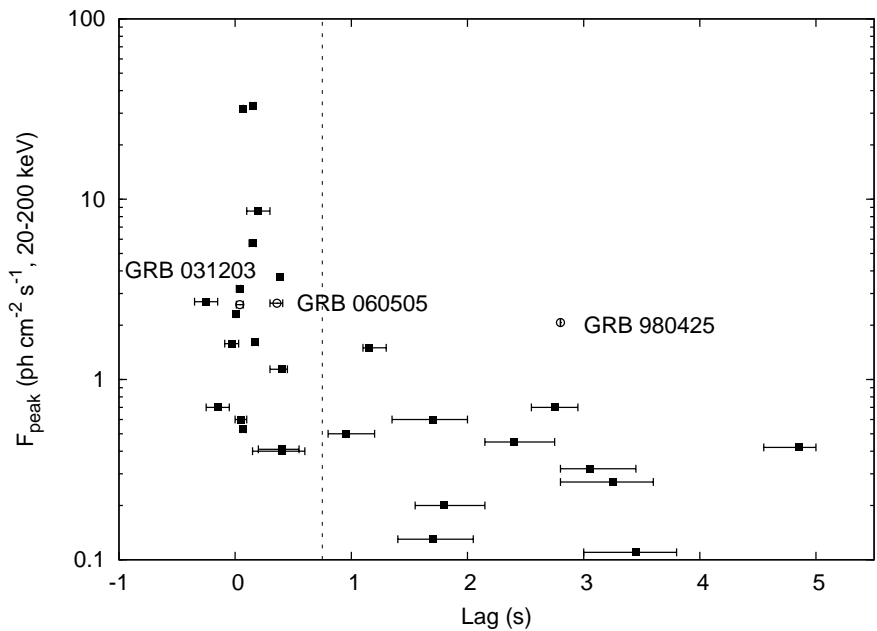}}
\caption{Spectral lag distribution of \textit{INTEGRAL} GRBs
  as a function of peak flux (20--200\,keV) \cite{foley:2008}. The SN bursts GRB\,980425 and GRB\,031203 are identified and represented by open circles, as is GRB\,060505 which does not have an associated SN. The SN burst XRF\,060218 has a peak flux of $0.6\,\rm{ph}\,\rm{cm}^{-2}\,\rm{s}^{-1}$ and is not included in the figure because of its very long lag of $61\pm26$\,s. GRBs with the
  longest lags tend to have low peak fluxes, whereas GRBs with short
  lags have both low and high peak fluxes. The dashed line indicates
  the separation between long and short-lag GRBs at $\tau=0.75$\,s.}
\label{fig:fp_lag}
\end{figure}

One notable feature of time profiles of long GRBs is the tendency for emission in the high energy band to lead the arrival of photons in a low energy band \cite[e.g.][]{norris:2002,foley:2008}. The observed lag of a GRB is a direct consequence of its spectral evolution because the peak of the $\nu\,F_{\nu}$ spectrum decays during the burst and hence allows the temporal and spectral properties to be combined in a single measurement. The spectral lag versus peak flux distribution of INTEGRAL GRBs is given in Fig.~\ref{fig:fp_lag} including the supernova (SN) bursts GRB\,980425 and GRB\,031203. The figure shows that both bright and faint GRBs have short spectral lags but there is an obvious absence of bright long-lag GRBs. These results are in good agreement with BATSE \cite{norris:2002} where the proportion of long-lag bursts increases at lower values of the peak flux.

At least some of the long-lag GRBs appear to belong to a sub-class of long bursts that are at closer distances and have a probable association with the supergalactic plane \cite{norris:2002,foley:2008}. More redshift measurements of faint GRBs are needed to establish the nature of this population. The spectral lag has been anti-correlated with the isotropic peak luminosity \cite{norris:2000} and is one of several correlations that have been discovered in the properties of the prompt emission of GRBs \cite[e.g.][]{amati:2006}. The correlations are very interesting but it is still most important to have the direct measurement of the redshift of the GRB.


\section{Polarisation}

The emission mechanism and magnetic fields in GRB jets are highly uncertain. There are three prominent emission mechanisms that include synchrotron emission with a small scale random magnetic field, synchrotron emission in a globally ordered magnetic field and the Compton drag model. The latter two mechanisms can yield high levels of polarisation. Polarisation measurements of a number of GRBs will be able to distinguish between the models \cite{toma:2009}.

\begin{figure}
\begin{center}
\mbox{
\subfigure{\includegraphics[width=0.5\columnwidth,height=0.3\textheight]{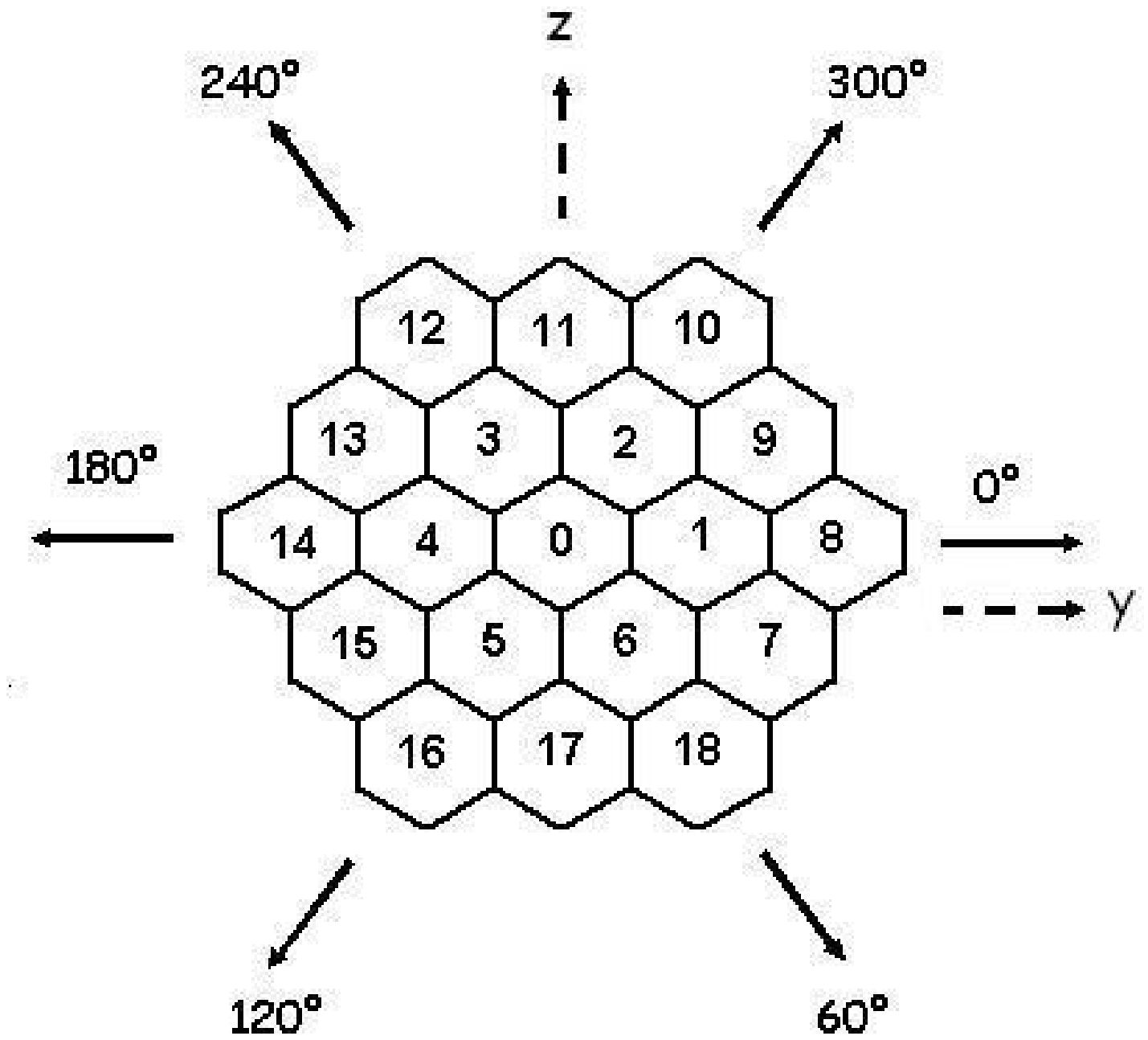}}
\subfigure{\includegraphics[width=0.5\columnwidth,height=0.3\textheight]{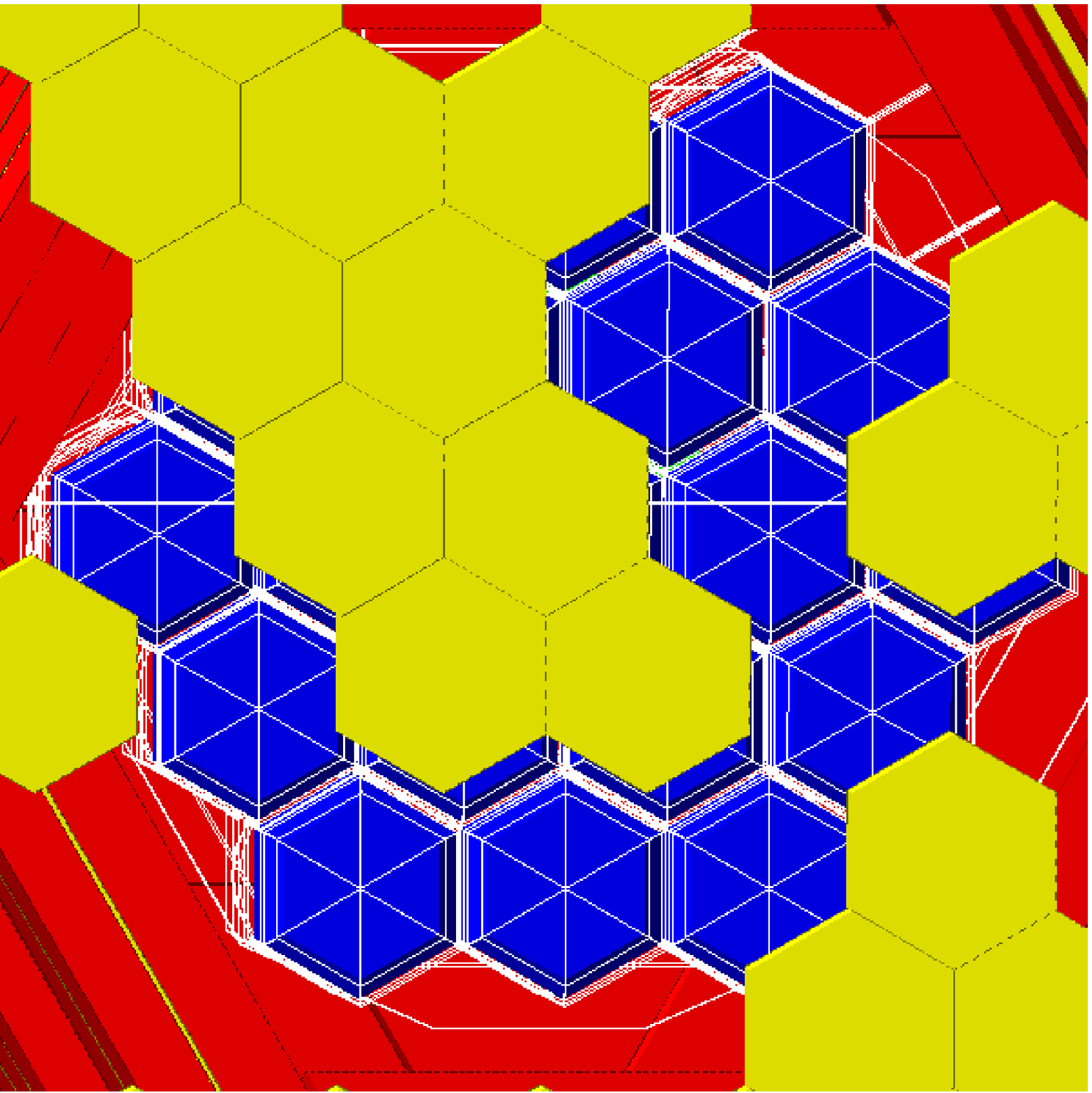}}}
\caption{Shown on the left is the numbering system used for the 19 SPI Germanium detectors and the 6 azimuthal directions used in the polarisation analysis. The figure on the right shows the projection or shadowgram of the coded mask elements (yellow) onto the 19 Germanium detectors (blue), as viewed from the direction of the
  incoming GRB photons generated from the GEANT 4 simulations for GRB\,041219A \citep{mcglynn:2007}. }
\label{mask_sim}
\end{center}
\end{figure}

The dominant method for detecting gamma-rays in the energy range of a few hundred keV is Compton scattering. Linearly polarised gamma-rays preferentially scatter perpendicular to the incident polarisation vector resulting in an azimuthal scatter angle distribution which is modulated relative to the distribution of unpolarised photons. The spectrometer SPI and imager IBIS on \textit{INTEGRAL} are not optimised to act as polarimeters but do respond because of the detector layout (Fig.~\ref{mask_sim}). The polarisation can be measured through multiple scatter events involving two or more detectors. GRB\,041219A is the brightest GRB detected by \textit{INTEGRAL} \cite{mcbreen:2006} and high levels of polarisation were measured in the brightest 12 seconds of this  GRB with SPI \cite{mcglynn:2007,mcglynn:2009} and varying polarisation with IBIS \cite{gotz:2009}. A high level of optical polarisation of 10$\%$ was recently measured in the afterglow of GRB\,090102 using the RINGO polarimeter on the Liverpool Telescope \cite{steele:2009}. In this context it is interesting to note that the absence of bright optical afterglows in most GRBs can be explained by a high magnetic field in the jet \cite[e.g.][]{gomboc:2009,zhang:2004}. This field is only beginning and there is a need for many more measurements of the polarisation of the prompt and the afterglow emission of GRBs from the optical through to $\gamma$-rays. 

\section{Faint GRBs}

\begin{figure}
\resizebox{\textwidth}{0.38\textheight}{\includegraphics{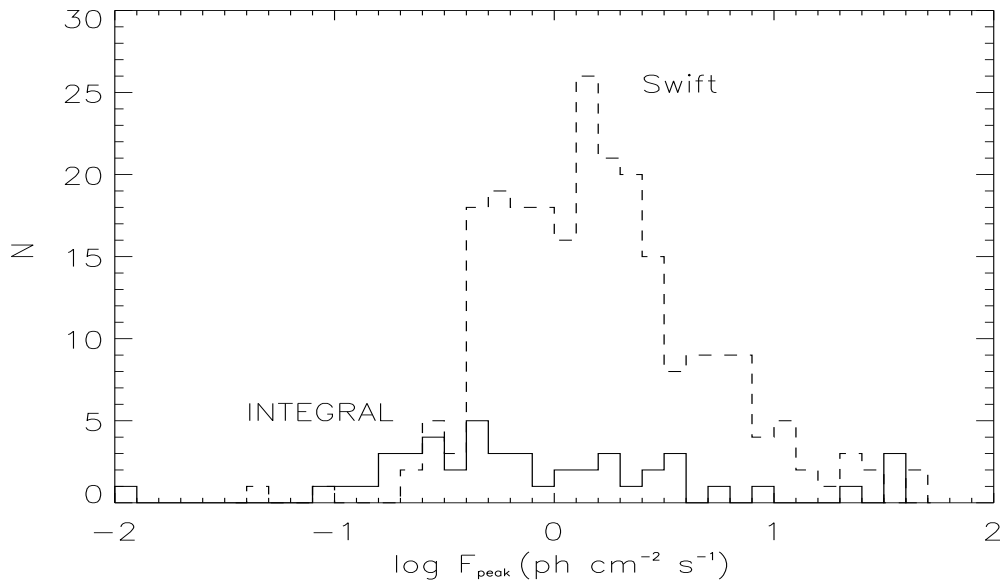}}
\caption{Peak flux distribution for GRBs detected by
  \textit{INTEGRAL} (20--200\,keV, solid line) and \textit{Swift}
  (15--150\,keV, dashed line). The \textit{Swift} data for 237 GRBs is taken from
http://swift.gsfc.nasa.gov/docs/swift/archive/grb\_table.html.}
\label{fig:fpeak}
\end{figure}

The peak flux (20--200\,keV) distribution of GRBs detected by IBIS and the peak flux (15--150\,keV) distribution of GRBs by the BAT instrument on \textit{Swift} is shown in Fig.~\ref{fig:fpeak} \cite{foley:2008}. IBIS detects proportionaly more faint GRBs than \textit{Swift} because of its better sensitivity in a field of view that is smaller by a factor of $\sim$12. The IBIS all-sky rate of GRBs is $\sim$1400\,yr$^{-1}$ in the energy range of 20--200\,keV above a peak flux of 0.15\,photons\,cm$^{-2}$\,s$^{-1}$ \cite{vianello:2009}. 

\section{High Energy Emission}

There are two misions in orbit, \textit{AGILE} launched in 2007 and \textit{Fermi} launched in 2008, that cover the X-ray and high energy gamma-ray bands and provide simultaneous broadband coverage. The number of GRBs detected above 100\,MeV is 15 and they reveal a wide variety of behaviour which is at least consistent with the GRB phenomenon. The main results are presented in Omodei et al. (2009) \cite{omodei:2009} and include:

1) The emission above 100\,MeV frequently starts later than the low energy emission and continues long after it, e.g. GRB\,080916C which is the most energetic burst at $\sim9\times10^{54}$\,ergs \cite{abdo:2009a}.

2) GRB\,080916C is fit over the energy from 8\,keV to 12\,GeV with a Band spectrum \cite{abdo:2009a} whereas GRB\,090902B requires a new power law component at low and high energies in addition to the usual Band function \cite{abdo:2009b}.

3) The short bursts GRB\,081024B and GRB\,090510 has delayed high energy emission \cite{giuliani:2009}. It is clear that there is a whole range of new phenomena to be discovered and studied in the high-energy $\gamma$-ray domain.

\section{X-ray and Optical Afterglows and Host Galaxies}

\begin{figure}
\mbox{
\subfigure{\resizebox{0.45\textwidth}{0.38\textheight}{\includegraphics{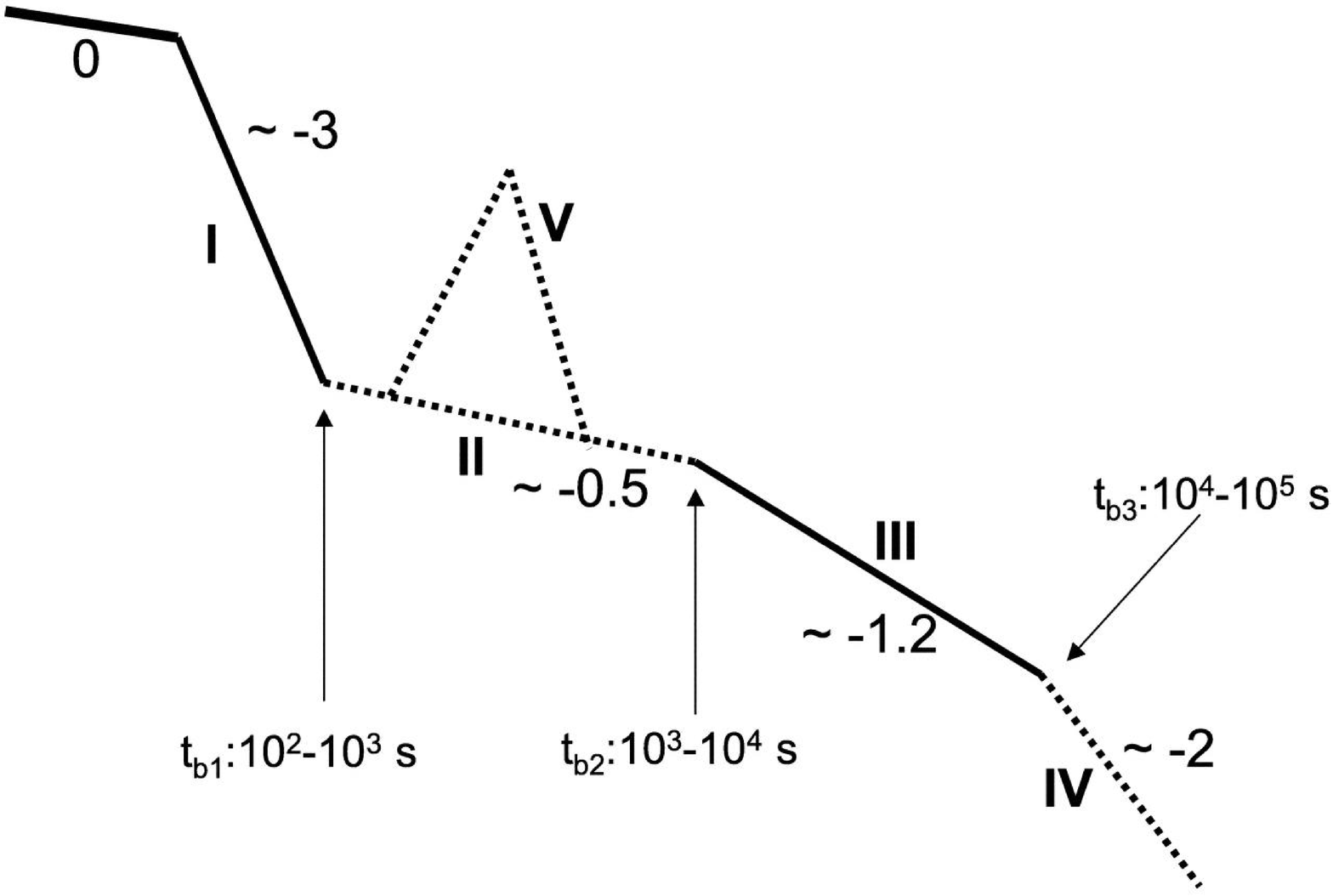}}}
\subfigure{\resizebox{0.45\textwidth}{0.38\textheight}{\includegraphics[bb=0 0 373 368, clip=]{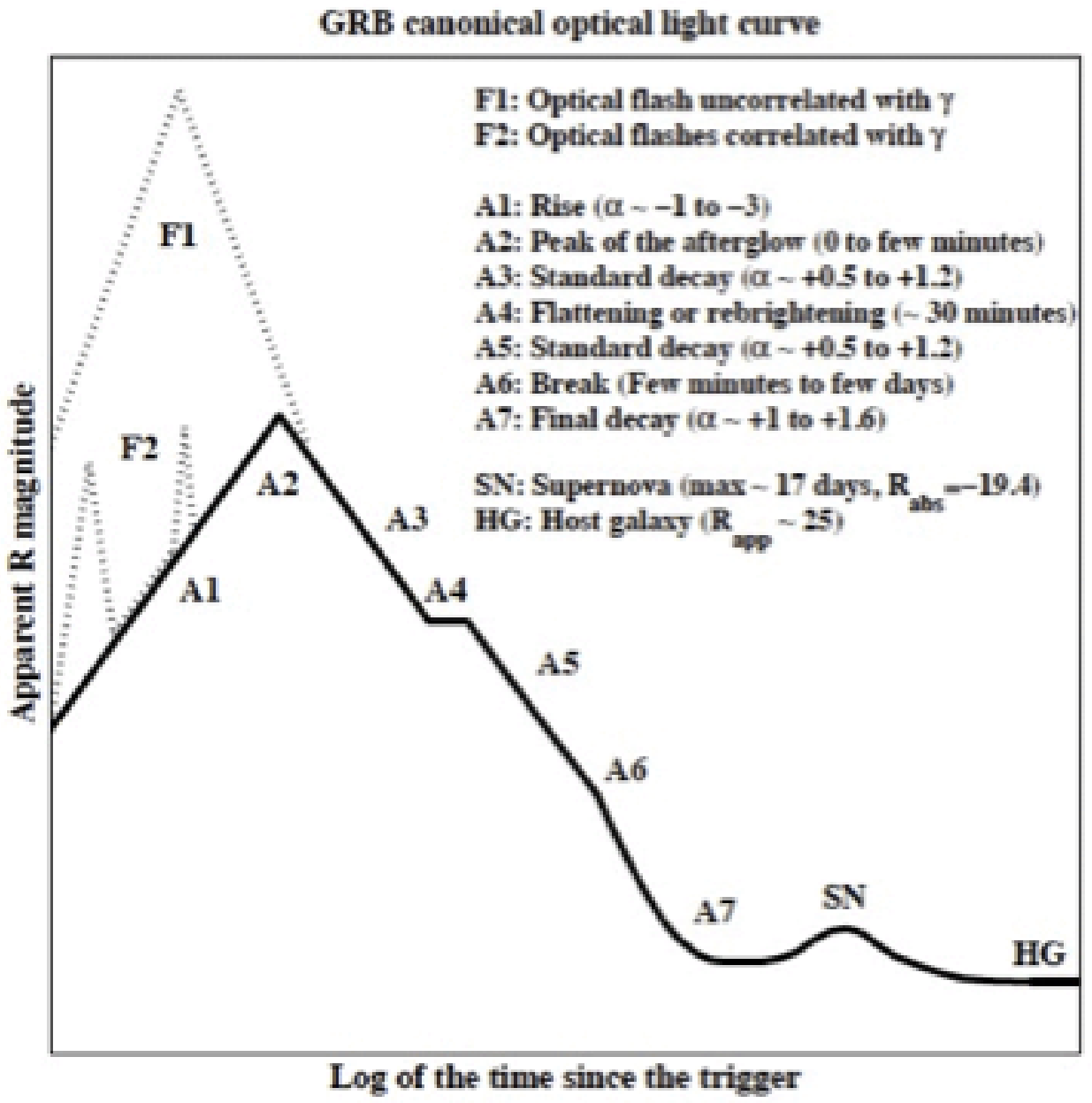}}}}
\caption{Left: the canonical X-ray afterglow showing the 5 main regions that are observed in most X-ray afterglows \cite{zhang:2006}. Right: the canonical optical afterglow showing the main features that are observed in most GRBs \cite{gendre:2009}.}
\label{fig:lcs}
\end{figure}

\textit{Swift} has filled in the temporal gap between the prompt emission and the afterglow for a very large number of bursts. The canonical X-ray afterglow (Fig.~\ref{fig:lcs}) has five main features. Initially there is a steep decline (I) followed by a plateau or shallow slope phase (II) between $10^{2}$--$10^{3}$ seconds and then the classical afterglow (III) followed by a jet break at $10^{5}$--$10^{6}$\,s (IV). In addition about half of the afterglows have superimposed X-ray flares (V) that are probably caused by delayed action in the central engine. The initial spectral break, between I and II, occurs when the steeply decaying tail of the emission from the prompt phase at large angles is overtaken by the slowly decaying emission from the forward shock.

The results from \textit{Swift} and an armada of robotic and large telescopes are slowly revealing the properties of the optical afterglow. The main properties of the optical afterglow of many bursts are briefly summarised in a canonical curve (Fig.~\ref{fig:lcs}). The very bright optical emission (F1) that occurs in some GRBs is often attributed to the reverse shock. If this emission is not dominant then the prompt optical emission accompanying the gamma-rays can be detected (F2). The optical afterglow (A1) rises and peaks after a few hundred seconds (A2) and is followed by the standard afterglow decay (A3 and A5) that can be interrupted by a flattening phase (A4). After the expected spectral jet break (A6) the supernova (SN) and host galaxy (HG) may be detectable. 

The host galaxies of about 100 long duration GRBs have been observed with HST and ground-based telescopes over a wide range in z. The general result is that GRB hosts are sub-L$^*$ galaxies with exponential disk profiles and high star formation rates on a per unit mass basis. The GRB selected hosts are free from selection effects and are powerful probes of galaxies and the Universe. The metallicities of GRB host galaxies at $z<1$ are sub-solar with a typical value of $\sim$0.1 \cite{gehrels:2009}.

%







\end{document}